\documentclass[floatfix,aps,twocolumn,showpacs,preprintnumbers,amsmath,amssymb]{revtex4-1}

\usepackage[applemac]{inputenc}
\usepackage[draft]{hyperref}
\usepackage{float}
\usepackage{amsfonts,dsfont}
\usepackage{graphicx}
\usepackage{listings}
\usepackage{ulem}
\usepackage{bm}
\usepackage{bbold}
\usepackage{braket}
\usepackage{todonotes}
\usepackage{url}
\usepackage{mathtools}

\newcommand{\bF}{\mathbf{F}}
\newcommand{\PI}{\mathbf{\Pi}}
\newcommand{\bPhi}{\mathbf{\Phi}}
\newcommand{\fo}{\mathbf{f}_0}

\newcommand{\bE}{\mathbf{E}}
\newcommand{\bcalE}{\boldsymbol{\mathcal{E}}}
\newcommand{\bA}{\mathbf{A}}
\newcommand{\bB}{\mathbf{B}}

\newcommand{\bH}{\mathbf{H}}
\newcommand{\bcalH}{\boldsymbol{\mathcal{H}}}
\newcommand{\bS}{\mathbf{S}}
\newcommand{\bGamma}{\mathbf{\Gamma}}

\newcommand{\br}{\mathbf{r}}

\newcommand{\bp}{\mathbf{p}}

\newcommand{\bcalP}{\boldsymbol{\mathcal{P}}}
\newcommand{\bcalM}{\boldsymbol{\mathcal{M}}}

\newcommand{\tk}{\tilde{k}}
\newcommand{\tq}{\tilde{q}}

\def\eps{\varepsilon}
\def\Re{\mathrm{Re}}
\def\Im{\mathrm{Im}}

\DeclareTextSymbol{\degre}{OT1}{23}

\begin{document}

\title{Force and torque on an electric dipole by spinning light fields}

\author{Antoine Canaguier-Durand}
\affiliation{ISIS \& icFRC, University of Strasbourg and CNRS, 8 all\'{e}e Gaspard Monge, 67000 Strasbourg, France.}
\author{Aur\'{e}lien Cuche}
\altaffiliation{Present address: CEMES, University of Toulouse and CNRS (UPR 8011), 29 rue Jeanne Marvig, BP 94347, 31055 Toulouse, France.}
\affiliation{ISIS \& icFRC, University of Strasbourg and CNRS, 8 all\'{e}e Gaspard Monge, 67000 Strasbourg, France.}
\author{Cyriaque Genet}
\email[Corresponding author:]{genet@unistra.fr}
\affiliation{ISIS \& icFRC, University of Strasbourg and CNRS, 8 all\'{e}e Gaspard Monge, 67000 Strasbourg, France.}
\author{Thomas W. Ebbesen}
\affiliation{ISIS \& icFRC, University of Strasbourg and CNRS, 8 all\'{e}e Gaspard Monge, 67000 Strasbourg, France.}

\begin{abstract}
We calculate the optical force and torque applied to an electric dipole by a spinning light field. We find that the dissipative part of the force depends on the orbital energy flow of the field only, because the latter is related to the  phase gradient generalized for such a light field. As for the remaining spin energy flow, it gives rise to an optical torque. 
The resulting change in the optical force is detailed for different experimentally relevant configurations, and we show in particular how this change is critical when surface plasmon modes are involved. 
\end{abstract}

\maketitle

\section{Introduction}
Direct manipulation of particles through light-induced forces has led to formidable progress which has been impacting research in many areas ranging from ultra cold matter physics to biology \cite{lewenstein2007ultracold,moffitt2008recent}. For the smallest objects, that can be handled in the Rayleigh regime, the optical forces induced by simple propagating laser beams are usually separated into two components: gradient forces directed towards the regions of highest field intensities, and radiation pressures directed along the Poynting vector \cite{stenholm1986semiclassical,cohen1992atom}. 

The rise of nano-optics has offered the experimentalists new types of optical excitations associated with inhomogeneous fields and complex beam topologies \cite{dogariu2012optically}.  Among these, surface plasmon (SP) modes have revealed themselves particularly efficient in trapping \cite{juan2011plasmon}, propelling \cite{wang2009propulsion,cuche2012brownian}, and sorting \cite{cuche2013sorting} nanoparticles, all in a great variety of environments, with for example recent implications in microfluidics \cite{kim2012joining} and atomic physics \cite{PhysRevLett.109.235309}. 

It was pointed out recently that the radiation pressure applied on an electric dipole by an inhomogeneously spinning light field is not given by the Poynting vector {\cite{arias2003optical,wong2006gradient,albaladejo2009scattering,berry2009optical,bekshaev2013subwavelength,bliokh2013dual}. In this article, a generalization of the phase gradient for a general harmonic field allows us to demonstrate that the radiation pressure is determined by the sole orbital part of the Poynting vector, with no contribution from its spin part. This has important consequences and we study how the modification of the Poynting vector can affect experiments involving optical forces generated by an evanescent field, such as total internal reflection or surface plasmons.  

\section{Force and torque on an electric dipole}
The Lorentz law gives the instantaneous force exerted on an electric dipole by general real electromagnetic fields $\left( \bcalE, \bcalH \right)$, supplemented  by the torque acting on the dipole \cite{stenholm1986semiclassical}:
\begin{align}\label{force_inst_real}
&\bF = \left( \bcalP \cdot \nabla \right) \bcalE + \mu_0 \dot{\bcalP} \times \bcalH
&\bGamma = \bcalP \times \bcalE ~ .
\end{align}
We assume in this work that the electromagnetic fields are monochromatic with an angular frequency $\omega$. They thus write in complex notations as $\bcalE=\Re(\bE)$ and $\bcalH=\Re(\bH)$ with $\bE(\br,t) = \bE_0 (\br) e^{-\imath \omega t}$ and $\bH(\br,t) = \bH_0 (\br) e^{-\imath \omega t}$. The electric dipole $\bcalP=\Re(\bp)$ is initially fixed immobile at position $\br$ in a medium of homogeneous and real refractive index $n(\omega)$. The complex dipolar moment $\bp = \bp_0(\br) e^{-\imath \omega t}$ is related to the electric field through an isotropic complex polarizability $\alpha (\omega)$ with $\bp_0(\br) = n^2 \alpha \bE_0(\br)$ in SI units \cite{FOOTNOTE}. 

Substituting complex fields and dipolar moment into Eq.~(\ref{force_inst_real}) leads to the time-averaged force~\cite{stenholm1986semiclassical,chaumet2000time}
\begin{align}
\left< \bF \right>_T  
&=\frac{n^2}{2} \Re \left[ \alpha \fo \right] ~ , \label{alpha0_f0} \\ 
  \mbox{ with } ~ ~   \fo &= \left( \bE_0 \cdot \nabla \right) \bE_0^{*} - \imath \mu_0 \omega  \bE_0 \times \bH_0^{*}  \label{f0_EH} \\
&= \left( \begin{array}{c} E_x \partial_x E_x^{*} + E_y \partial_x E_y^{*} + E_z \partial_x E_z^{*} \\ \vdots \end{array} \right) ~ . \label{f0_compact}
\end{align}
Eq.~(\ref{alpha0_f0}) can be used to perform the usual decomposition of the force into reactive and dissipative components, yielding the gradient force and radiation pressure, respectively \cite{cohen1992atom}. The reactive force, proportional to $\Re[\fo]$,  is easily obtained from Eq.~(\ref{f0_compact}) and can be interpreted as an intensity gradient so that
\begin{align} \label{F_reactive}
\bF_\mathrm{reactive} 
= \frac{n^2}{4}  \Re[\alpha] \nabla \left( \left\| \bE_0 \right\|^2 \right) ~ .
\end{align}
The dissipative force, proportional to $\Im[\fo]$, writes from Eq.~(\ref{f0_EH}) after some algebra as
\begin{align} \label{F_dissipative}
\bF_\mathrm{dissipative} & 
= n^2 \omega \mu_0 \Im[\alpha] \left( \PI - \frac{\nabla \times \bPhi_E}{2\omega \mu_0}\right)~, 
\end{align}
where $\PI = \frac{1}{2} \Re \left[ \bE_0 \times \bH_0^{*}\right] = \left< \bcalE \times \bcalH \right>_T$ is the time-averaged Poynting vector and $\bPhi_E = -\Im[\bE_0 \times \bE_0^{*}] / 2 = \bcalE \times \dot{\bcalE} /\omega$ the time-independent electric polarization ellipticity. 

This derivation gives the crucial result that the radiation pressure exerted on an electric dipole is not proportional to the Poynting vector as soon as the ellipticity of the acting field has a non-vanishing curl. This observation has led to interpreting the curl term in Eq.~(\ref{F_dissipative}) as a third force component associated with the spin density of the field~\cite{arias2003optical,wong2006gradient,albaladejo2009scattering,nieto2010optical,gomez2011nonconservative,wang2012optical,iglesias2012light,MarquezOptLett2012,ruffner2012optical}. But as we now show through a generalization of the phase gradient, the spin part of the Poynting vector does not play a role in the radiation force. We emphasize that this conclusion is reached from the same mathematical quantities as those used in \cite{arias2003optical,wong2006gradient,albaladejo2009scattering,nieto2010optical,gomez2011nonconservative,wang2012optical,iglesias2012light,ruffner2012optical}.

\section{Physical interpretation of the force}
Eq.~(\ref{F_dissipative}) gives a decomposition of the time-averaged Poynting vector $2\omega\mu_0\PI=-\Im[\fo]+\nabla\times\bPhi_E$ which actually corresponds to separating $\PI$ into its orbital and spin parts with respect to the electric field -superscript $(E)$. To explain this point, let us first write the complex polarization of the electric field as $\bE_0(\br)=\bA(\br)+\imath \bB(\br)$ with real vectors $\bA(\br),\bB(\br)$. One can show, following \cite{berry2001polarization}, that $\bPhi_E= \bA \times \bB$ is normal and proportional to the surface of the ellipse formed by the electric field over a time period $\frac{2\pi}{\omega}$. This justifies the interpretation of $\bPhi_E$ as the direction and magnitude of ellipticity of the electric field, also related to the electric chirality flow of the field \cite{bliokh2012transverse}. 

Moreover, $\bPhi_E$ is proportional to the local expectation value $\bS$ for the spin operator of the field with $\bPhi_E ~=~ \left< \left\| \bcalE \right\|^2 \right>_T \bS$~\cite{berry2001polarization}. The curl component of the time-averaged Poynting vector can thus be written as
\begin{align}\label{Phi_spin}
\frac{\nabla \times \bPhi_E}{2\omega \mu_0}=\frac{\| \bE_0 \|^2 }{4 \omega \mu_0 } \nabla \times \bS + \frac{1}{4 \omega \mu_0 } \nabla \left( \| \bE_0 \|^2 \right) \times \bS
\end{align}
and is identified with the spin part. Then the decomposition of $\PI$ into orbital and spin parts can be written as:
\begin{align}\label{spin_orbital}
&\PI_O^{(E)} = -\frac{\Im[\fo]}{2\omega \mu_0}
&\PI_S^{(E)} = \frac{\nabla \times \bPhi_E}{2\omega \mu_0} ~ ,
\end{align}
in agreement with~\cite{berry2009optical}. The connection with the spin operator $\bS$ drawn in Eq.~(\ref{Phi_spin}) is important as it shows that the field must have either an inhomogeneous spin or a non-zero spin with an inhomogeneous intensity in order to have $\PI_S^{(E)}$ different from zero. When neither of those conditions are met, $\PI \equiv \PI_O^{(E)}$ and the usual form of the radiation pressure is recovered. 

Finally, this decomposition shows that subtracting the curl term to the time-averaged Poynting vector in the dissipative force only leaves the orbital energy flow to contribute to the radiation pressure \cite{berry2009optical,bekshaev2013subwavelength,bliokh2013dual}:
\begin{align}\label{F_dipole}
\left< \bF \right>_T &= \frac{n^2}{4} \Re[\alpha] \nabla \left( \left\| \bE_0 \right\|^2 \right) + n^2  \omega \mu_0  \Im[\alpha] ~ \PI_O^{(E)}~.
\end{align}
This statement will be supported below by interpreting, in specific cases, the orbital component as a phase gradient. 

Let us first recall that for electric fields with linear polarization $\bE_0(\br)=E_0(\br) \hat{\mathbf{y}}$, with a scalar component written as $E_0(\br)=\rho(\br) e^{\imath \phi(\br)}$, the term $\fo$ in Eq.~(\ref{alpha0_f0}) reduces to
\begin{align}\label{fo_gradients}
\fo = \rho \nabla \rho - \imath \rho^2 \nabla \phi
\end{align}
and yields the usual interpretation of the reactive $\Re[\fo]$ and dissipative $\Im[\fo]$ components of the force proportional to amplitude and phase gradients respectively \cite{cohen1992atom}. 

In the general polarization case however, each component $E_0^{j}=\rho^j e^{\imath\phi^j}$ has its own amplitude and phase. From Eq.~(\ref{f0_compact}), $\Re[\fo]$ can still be written as an amplitude gradient, hence the expression for the reactive force in Eq.~(\ref{F_reactive}). In contrast, $\Im[\fo]=- \sum_j (\rho_j)^{2}\nabla \phi_j$ is a weighted average of the phase gradients of each component involved. This expression can be seen as a generalization of the phase gradient. Indeed, when the additional assumption is made that the three phase gradients are identical $(\nabla \phi_x=\nabla \phi_y = \nabla \phi_z =: \nabla \phi)$, the usual expression (\ref{fo_gradients}) is recovered, showing therefore the fundamental connection between the phase gradient and the orbital component of the Poynting vector. In this sense, $\PI_O^{(E)}=\PI-(\nabla \times \bPhi_E) / (2\omega \mu_0)$ has to be seen as a modified Poynting vector in the transfer of electromagnetic energy. Noteworthy, this modification operated via a curl term still satisfies $\nabla\cdot \PI_O^{(E)} =\nabla\cdot \PI$ and therefore amounts to a different choice of gauge that does not affect Poynting's theorem.

\section{Physical interpretation of the torque}
Meanwhile, the time-independent torque turns out to be directly proportional to the field ellipticity with
\begin{align}\label{T_dipole}
\bGamma &= n^2 \Im[\alpha] \bPhi_E ~,
\end{align}
as suggested in \cite{bliokh2013dual}. To clarify its physical meaning, one first notes that, when time evolves, the dipolar moment $\bcalP(\br)$ rotates in the same plane as the electric field $\bcalE(\br)$, normal to the ellipticity: $\bcalP \times \dot{\bcalP} = \left| n^2 \alpha \right|^2 \bcalE \times \dot{\bcalE} = \omega  \left| n^2 \alpha \right|^2 \bPhi_E$. The non-zero value of the vector product $\bGamma=\bcalP\times \bcalE$ then exhibits a phase-lag between the source $\bcalE$ and the linear response $\bcalP$. With $\Im[\alpha]\geq 0$, $\bcalP$ is delayed with respect to $\bcalE$. Similarly to the case of a driven damped harmonic oscillator, this delay is due to dissipation in direct relation with the factor $\Im[\alpha]$ in Eq.~(\ref{T_dipole}). The torque thus works towards aligning $\bcalP$ to $\bcalE$, trying to follow the field source. The amount of energy given away by the torque to the dipole is simultaneously lost through dissipation (heat).

We emphasize how Eqs.~(\ref{F_dipole}) and (\ref{T_dipole}) display a remarkable balance between the radiation pressure and the orbital energy flow on the one hand, and the exerted torque and the spin energy flow on the other hand. We conclude that the incoming electromagnetic field transfers mechanical energy to the electric dipole through dissipation via two different channels: after time-averaging, the orbital part $\PI_O^{(E)}$ gives rise to a dissipative net force (i.e. radiation pressure) while the spin part $\PI_S^{(E)}$ can be related to the torque applied to the dipole to maintain it rotating with the electric field. 

We finally note that the decomposition in orbital and spin parts for the Poynting vector is asymmetrically driven by the electric field with $\PI=\PI_O^{(E)}+\PI_S^{(E)}$ because our model of an electric dipole only reacts to $\bcalE$. Considering a magnetic dipole $\bcalM=\Re[\beta \bH]$ gives, by symmetry, a similar result with an $\bcalH$-driven $\PI=\PI_O^{(H)}+\PI_S^{(H)}$ decomposition~\cite{bekshaev2013subwavelength,bliokh2013dual}.

We now describe a few specific field distributions that enable to illustrate this discussion most appropriately. All the cases discussed below are associated with transverse magnetic (TM) polarized waves, either in the near or far field.

\section{Evanescent TM-polarized waves}
We start from the cartesian general expression for a TM-polarized evanescent wave, invariant in the $y$-direction with
\begin{align*}
&\bE_0 =  E_0 ~ e^{\imath k x} e^{\imath q z} \left( \tq , 0, -\tk \right)^t \\
&\bH_0 =  \sqrt{n^2 \eps_0 / \mu_0} ~ E_0 ~ e^{\imath k x} e^{\imath q z} \left( 0, 1 , 0 \right)^t
\end{align*}
where the complex $k=k'+\imath k''$ and $q=q'+\imath q''$ fulfill $k^2 + q^2 = (\omega n/c)^2$ and the dimensionless $\tk= \frac{c}{n \omega} k$ and $\tq = \frac{c}{n \omega} q$ fulfill $\tk^2 + \tq^2 = 1$. We note that while the $x$ and $z$ components of $\bE_0$ have different phases, allowing for ellipticity, their phase gradients are identical with $\nabla \phi=\left( k' , 0 , q' \right)^t$. The associated time-averaged Poynting vector and spin expectation value are simply evaluated as \mbox{$\PI=\left(| E_0 |^2 / 2\omega \mu_0\right)e^{-2k''x-2q''z} \left( k' , 0 , q' \right)^t$} and 
\begin{align*}
\bS =2\frac{\tq'\tk''-\tk' \tq''}{ |\tk|^2+|\tq|^2} \left( 0, 1 , 0 \right)^t ~ .
\end{align*}%

The spin vector is homogeneous so that the second term in the right hand side of Eq.~(\ref{Phi_spin}) gives the only contribution to $\PI_S^{(E)}$. After some algebra, one can show that the orbital and spin components of the Poynting vector are collinear to it with $\PI_O^{(E)} =\left[ 1 +2 (\tk'')^2 + 2 (\tq'')^2 \right]  \PI$ and $\PI_S^{(E)} = -2 \left[ (\tk'')^2 + (\tq'')^2 \right] \PI$, meaning that $\PI$ and $\PI_S^{(E)}$ are in opposite directions. This collinearity allows us to define the relative difference $\Delta$ between the Poynting vector and its orbital component as $\PI_O^{(E)} - \PI= \Delta \PI$ with $\Delta = 2  \left[ (\tk'')^2 + (\tq'')^2 \right] \geq 0$ as a measure of the increase of the energy flow after this substitution. As seen on the latter equation, $\Delta$ stems from the evanescence of the field due to the homogeneous spin of such a field. 

The gradient force and radiation pressure of the TM-polarized field can be expressed as
\begin{align*}
\bF_\mathrm{reactive} &=  - \left( n^2 \left\| \bE_0 \right\|^2 / 2 \right) \Re[\alpha] \left( k'' , 0 , q''  \right)^t \\ 
\bF_\mathrm{dissipative} &= \left( n^2 \left\| \bE_0 \right\|^2 / 2 \right)  \Im[\alpha] \left( k' , 0 , q'  \right)^t
\end{align*}
with $ \left\| \bE_0 \right\|^2 =  \left| E_0 \right|^2 e^{-2k''x-2q''z} \left(1 + 2 (\tk'')^2 + 2(\tq'')^2\right) $. Given that $k ' k '' + q ' q '' = 0$, these two components are perpendicular to each other. Moreover, as all $E_0^{j}$ have the same phase gradient $\nabla \phi$, the two components of the force can be simply expressed in terms of the imaginary and real parts of the wave vectors. This directly supports the interpretation of the reactive and dissipative forces as amplitude and phase gradients, respectively.  

\section{Total internal reflection}
The phenomenon of total internal reflection (TIR) is described by such a TM-polarized evanescent field with real $\tk = n_1 / n_2 \sin\theta_1$ and imaginary $\tq = \imath \sqrt{\left( n_1/n_2\right)^2 \sin^2\theta_1 -1}$ components of the wave vector, given an incidence angle $\theta_1$ greater than the critical angle $\theta_C = \mathrm{arcsin} \left( n_2 / n_1\right)$ at the $(z=0)$ interface between two dielectrics of refractive indices $n_1 \geq n_2$. The relative change in the dissipative force then follows with $\Delta_\mathrm{TIR} = 2 \left( n_1/n_2\right)^2 \left( \sin^2\theta_1 - \sin^2\theta_C\right)$. 
\begin{figure}[htb]
\centering{
\includegraphics[width=6cm]{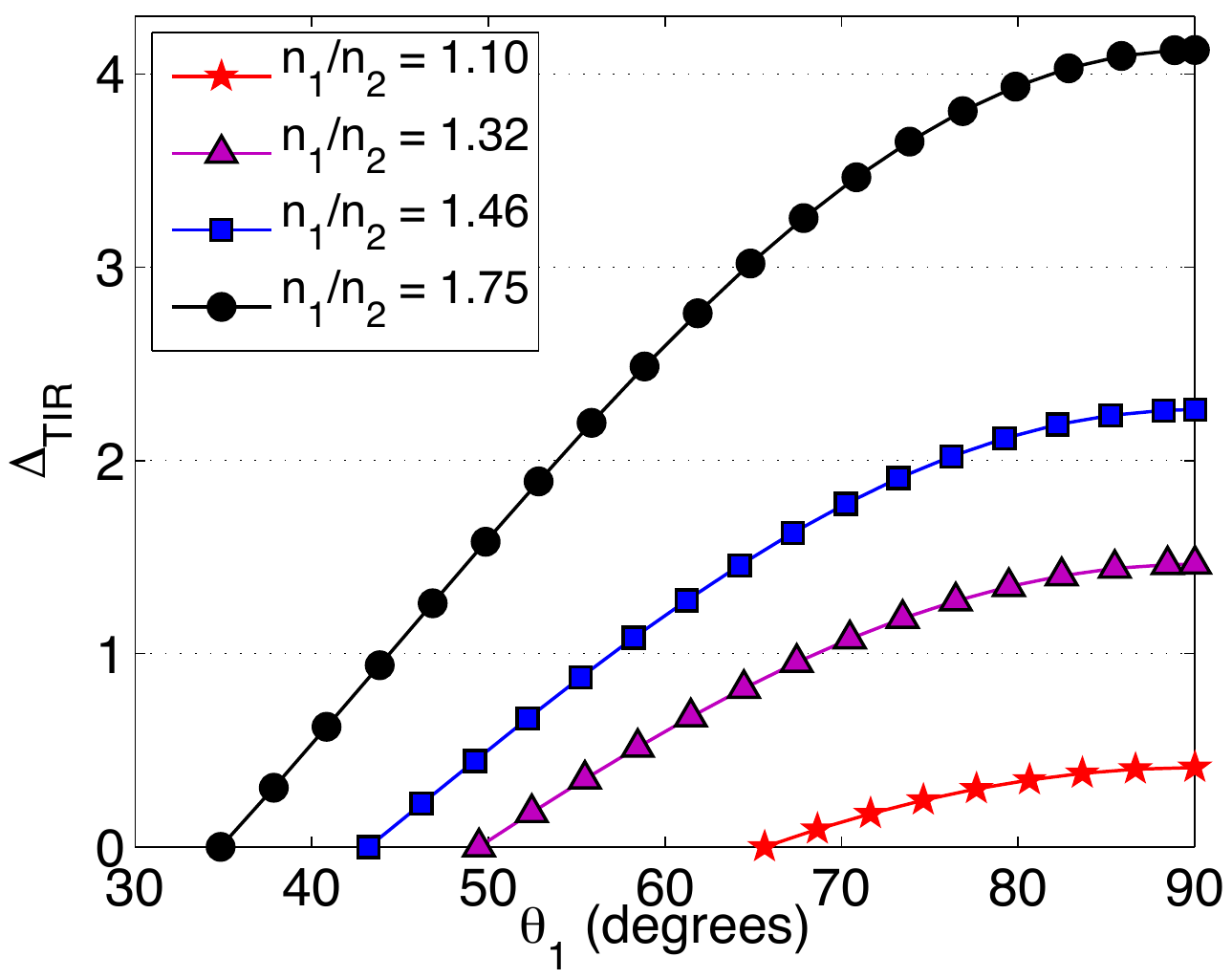}}
 \caption{(Color online) Relative change $\Delta_\mathrm{TIR}$, as a function of the incidence angle $\theta_1 \geq \theta_C$ for experimentally relevant values of the index contrast $n_1/n_2$ between the two dielectrics, with $n_1 \in \{ 1.46, 1.75\}$ for SiO$_2$ or quartz and $n_2 \in \{1, 1.33 \}$ for air or H$_2$O.}
 \label{Delta_TIR}
\end{figure}

This parameter $\Delta_\mathrm{TIR}$ is plotted in Fig.~\ref{Delta_TIR} as a function of the incident angle $\theta_1$ and it displays interesting features. For $\theta_1$ just above $\theta_C$, $\Delta_\mathrm{TIR}\simeq 0$ and the Poynting vector is equal to its orbital part. However, as the evanescence of the field in the $z$-direction increases with $\theta_1$, $\Delta_\mathrm{TIR}$ reaches non-negligible values, manifesting the onset of a spin contribution to the Poynting vector. 

\section{Surface plamon field} 
Such contribution is actually always present in the case of a plain SP field launched at an interface ($z=0$) between a metal and a dielectric, with dielectric functions $\eps_m(\omega)$ and $\eps_d(\omega)$. Here, the wave vector of the field is a complex quantity in both $x$ and $z$ directions with $\tk=  \sqrt{\eps_m/(\eps_d+ \eps_m)}$ and $\tq=  \sqrt{\eps_d/(\eps_d+ \eps_m)} $. It follows that the relative change in the dissipative force, due to the substitution of $\PI$ by its orbital part, is in this case frequency dependent with $\Delta_\mathrm{SP}~=~ 2 \left[ \Im \left(  \sqrt{\eps_m/(\eps_m+\eps_d)} \right) \right]^2  +2\left[ \Im \left( \sqrt{\eps_d/(\eps_m+\eps_d)} \right)  \right]^2$. The evaluation of this factor for a Au-H$_2$O interface is presented in Fig.~\ref{Delta_SP} as a function of the incident wavelength, together with the $y$-component $S_y$ of the spin, thereby stressing the relation. 
\begin{figure}[htb]
\centering{
\includegraphics[width=6cm]{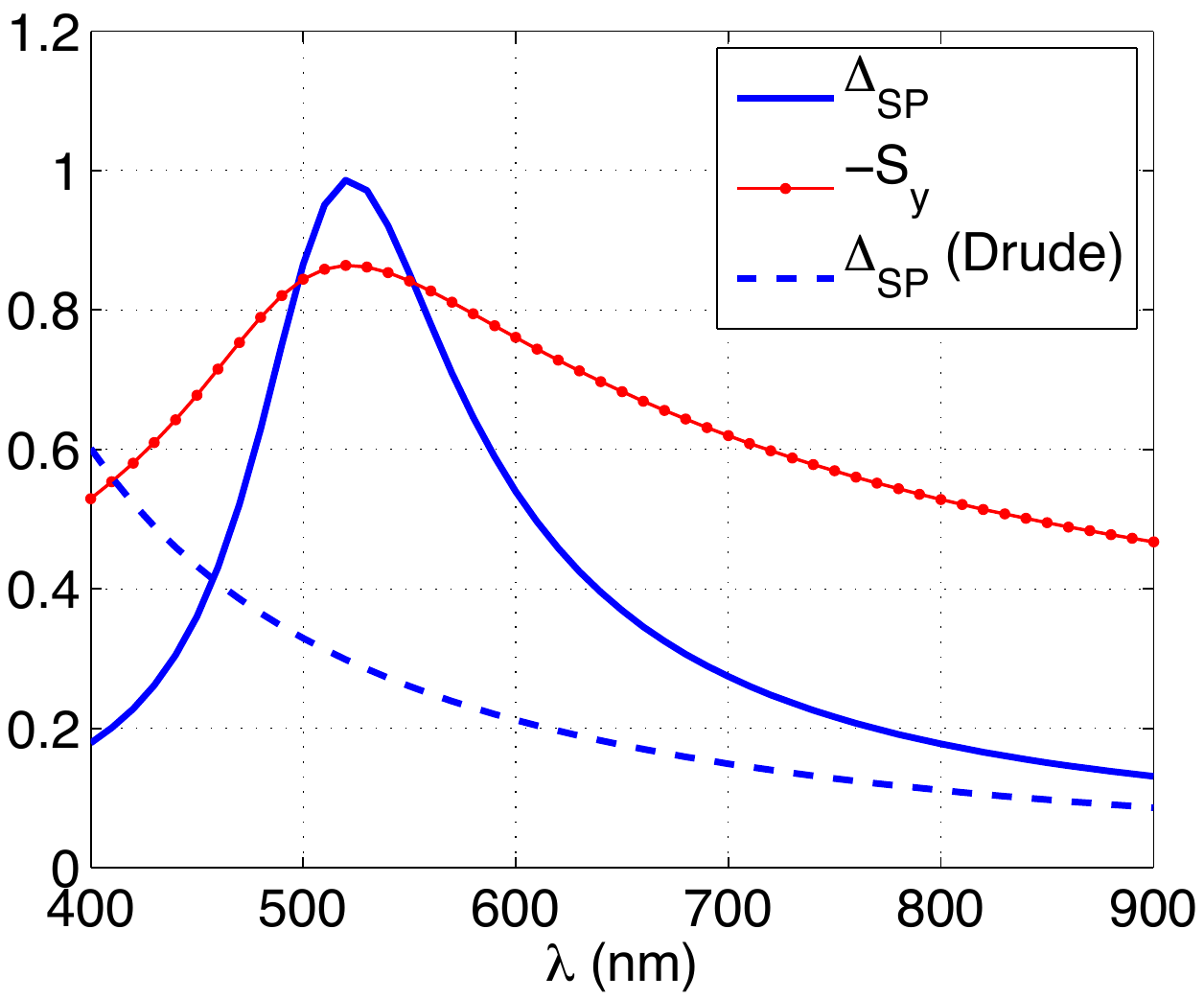}}
 \caption{(Color online) Relative change $\Delta_\mathrm{SP}$ for optical data (blue plain curve) and Drude model (blue dashed line) for a Au-H$_2$O $(n=1.33)$ interface. Remarkably, an evaluation based on a Drude model strongly depart from using optical data. The $y$-component of the spin is the red dotted curve, multiplied by $-1$. }
 \label{Delta_SP}
\end{figure}
We see that the Poynting vector always differs from its orbital part, in agreement with the intrinsic spinning nature of SP modes \cite{bliokh2012transverse}. The induced relative difference $\Delta_\mathrm{SP}$ for the dissipative force is maximal at $\lambda \simeq 520$ nm using optical data for gold. Interestingly, this points to the crucial role of the interband transitions in the generation of the ellipticity of the plasmonic field and stresses the importance of using optical data for realistic and trustful evaluations of plasmonic forces.

\section{Intersecting standing waves}
Finally, our different interpretation becomes totally clear when considering the situation of two TM-polarized intersecting standing waves (SW) addressed in \cite{albaladejo2009scattering}. We start from the fields 
\begin{align*}
\bE_\mathrm{sw} &=  E_\mathrm{sw}  \left( e^{\imath \varphi_0} \cos kz , 0 , -\cos kx \right)^t \\
\bH_\mathrm{sw} &=  \imath \sqrt{n^2 \eps_0 /\mu_0} E_\mathrm{sw} \left( \sin k x + e^{\imath \varphi_0} \sin k z \right) (0 , 1, 0)^t
\end{align*}
with real $k=\omega n/c $ and real phase shift $\varphi_0$ between the two standing waves. Here too, the two components of $\bE_\mathrm{sw}$ have different phases $\left(\phi_x\neq\phi_z\right)$ but their phase gradients both equal zero. As a consequence, the dissipative force, i.e. the radiation pressure, vanishes for this electromagnetic field and the optical force reduces to the gradient force. 

It is possible to obtain a field for which $\PI$ and $\PI_O^{(E)}$ are not colinear by merely adding a plane wave \mbox{$\bE_\mathrm{pw} =  E_\mathrm{pw}  e^{\imath k y}\left( 1 , 0 , 0\right)^t$} to the SW intersection. In this case, the generalized phase gradient $\nabla \phi$ induces a non-zero orbital Poynting vector in the $(y,z)$ plane with no component along $x$.  The evaluations are straightforward and reveal that in the $(x,z)$-plane of the two SWs, the dipole is only pushed by $\PI_O^{(E)}$ along the $z$ direction, while $\PI$ has a non-vanishing component along $x$. This constitutes thus a simple situation for which the radiation pressure is clearly directed off the orientation of $\PI$.
 
\section{Conclusion} 
To summarize, we stress that the optical force on an electric dipole is the sum of two terms only: the gradient force and the radiation pressure. The latter component is solely related to the orbital part of the Poyting vector of the driving field while the spin part is related to the optical torque exerted on the dipole. We have shown how the evanescent character of the TM-polarized field is directly related to the strength of this spin part, and in the case of SP fields, how the relative change in energy flow $\Delta_\mathrm{SP}$ between the Poynting vector and its orbital part depends strongly on the incident wavelength $\lambda$. Our work reveals how crucial it is to substitute properly this orbital part for the Poynting vector when expressing the dissipative force acting on the electric dipole. This has important consequences in the context of plasmonic manipulations of nanoparticles.

{\it Note added -}
Recently, Ruffner and Grier \cite{RuffnerCommentPRL2013} gave a similar expression of the radiation pressure as a phase gradient generalized to arbitrary polarization in a Comment to Ref. \cite{albaladejo2009scattering}. Like us, these authors emphasize that the curl of the spin angular momentum density does not contribute to the force experienced by a Rayleigh object.

\section{Acknowledgments} 
We acknowledge support from the ERC (Grant 227557) and from the French program Investissement d'Avenir (Equipex Union).


\begin{thebibliography}{99}

\bibitem{lewenstein2007ultracold} M. Lewenstein, A. Sanpera, V. Ahufinger, B. Damski, A. Sen, and U. Sen, Adv. Phys. {\bf 56}, 243 (2007).

\bibitem{moffitt2008recent} J. Moffitt, Y. Chemla, S. B. Smith, and C. Bustamante, Annu. Rev. Biochem. {\bf 77}, 205 (2008).

\bibitem{stenholm1986semiclassical} S. Stenholm, Rev. Mod. Phys. {\bf 58}, 699 (1986).

\bibitem{cohen1992atom} C. Cohen-Tannoudji, J. Dupont-Roc, and G. Grynberg, {\it Atom-photon interactions: basic processes and applications} (Wiley-Interscience, 1992).

\bibitem{dogariu2012optically} A. Dogariu, S. Sukhov, and J. J. S{\'a}enz, Nature Photon. {\bf 7}, 24 (2012).

\bibitem{juan2011plasmon} M. L. Juan, M. Righini, and R. Quidant, Nature Photon. {\bf 6}, 349 (2011).

\bibitem{wang2009propulsion} K. Wang, E. Schonbrun, and K. B. Crozier, Nano Lett. {\bf 7}, 2623 (2009).

\bibitem{cuche2012brownian} A. Cuche, B. Stein, A. Canaguier-Durand, E. Devaux, C. Genet, and T.W. Ebbesen, Nano Lett. {\bf 12} 4329 (2012).

\bibitem{cuche2013sorting} A. Cuche, A. Canaguier-Durand, E. Devaux, J. A. Hutchison, C. Genet, and T. W. Ebbesen, (submitted).

\bibitem{kim2012joining} J. Kim, Lab on a Chip {\bf 12}, 3611 (2012).

\bibitem{PhysRevLett.109.235309} M. Gullans, T. G. Tiecke, D. E. Chang, J. Feist, J. D. Thompson, J. I. Cirac, P. Zoller, and M. D. Lukin, Phys. Rev. Lett. {\bf 109}, 235309 (2012).

\bibitem{arias2003optical} J. R. Arias-Gonzalez and M. Nieto-Vesperinas, J. Opt. Soc. Am. A {\bf 20}, 1201 (2003).

\bibitem{wong2006gradient} V. Wong and M. A. Ratner, Phys. Rev. B {\bf 73}, 075416 (2006).
 
\bibitem{albaladejo2009scattering} S. Albaladejo, M. I. Marqu{\'e}s, M. Laroche, and J. J. S{\'a}enz, Phys. Rev. Lett. {\bf 102}, 113602 (2009).
  
\bibitem{berry2009optical} M. V. Berry, J. Opt. A {\bf 11}, 094001 (2009).

\bibitem{bekshaev2013subwavelength} A. Ya Bekshaev, J. Opt. {\bf 15}, 044004 (2013).

\bibitem{bliokh2013dual} K. Y. Bliokh, A. Ya Bekshaev, and F. Nori, New J. Phys. {\bf 15}, 033026 (2013).

\bibitem{FOOTNOTE} In order to fulfill the optical theorem, one usually replaces the dipole polarizability $\alpha _0$ by an effective polarizability $\alpha _\protect \mathrm  {eff}= \alpha _0 \left [ 1 - \imath \protect \frac  {2}{3} \protect \frac  {n^3 \omega ^3}{ c^3} \protect \frac  {n^2 \alpha _0}{4\pi \varepsilon _0} \right ]^{-1}$ which takes into account the field scattered by the dipole. Here we will simply use the notation $\alpha $ which can correspond to $\alpha _0$ or $\alpha _\protect \mathrm  {eff}$ depending on whether the radiated field of the dipole is included or not.

\bibitem{chaumet2000time} P. C. Chaumet and M. Nieto-Vesperinas, Opt. Lett. {\bf 25}, 1065 (2000).

\bibitem{nieto2010optical} M. Nieto-Vesperinas, J. J. S{\'a}enz, R. G{\'o}mez-Medina, and L. Chantada, Opt. Express {\bf 18}, 11428 (2010).

\bibitem{gomez2011nonconservative} R. G{\'o}mez-Medina, M. Nieto-Vesperinas, and J. J. S{\'a}enz, Phys. Rev. A {\bf 83}, 033825 (2011).
  
\bibitem{wang2012optical} L.-G. Wang,  Opt. Express {\bf 20}, 20814 (2012).

\bibitem{iglesias2012light} I. Iglesias and J. J. S{\'a}enz,  Opt. Express {\bf 20}, 2832 (2012).

\bibitem{MarquezOptLett2012} M. I. Marqu{\'e}s, M. Laroche, and J. J. S{\'a}enz, Opt. Lett. {\bf 37}, 2787 (2012).

\bibitem{ruffner2012optical} D. B. Ruffner and D. G. Grier, Phys. Rev. Lett. {\bf 108}, 173602 (2012).

\bibitem{berry2001polarization} M. V. Berry and M. R. Dennis, Proc. R. Soc. London A {\bf 457}, 141 (2001). Here we take $\bS$ as defined in Eq.~(3.2) in this reference, but multiplied by $(-1)$. 

\bibitem{bliokh2012transverse} K. Y. Bliokh and F. Nori, Phys. Rev. A {\bf 85}, 061801(R) (2012).

\bibitem{RuffnerCommentPRL2013} D. B. Ruffner and D. G. Grier, Phys. Rev. Lett. {\bf 111}, 059301 (2013).

\end{thebibliography}

\end{document}